\documentclass{article}
\usepackage{amsmath, amssymb, graphics, setspace}

\newcommand{\mathsym}[1]{{}}
\newcommand{\unicode}[1]{{}}

\usepackage{color}

\newtheorem{theorem}{Theorem}

\newtheorem{corollary}[theorem]{Corollary}
\newtheorem{proposition}[theorem]{Proposition}

\usepackage{authblk}

\begin{document}

\title{Jacobi Group Symmetry of Hamilton{'}s Mechanics}
\author[$\star$]{Stephen G. Low} 

\author[$\dagger$$\ddagger$]{Rutwig Campoamor-Stursberg} 
\affil[$\star$] {\small \textit{E-mail}: \texttt{Stephen.Low@utexas.edu} } 
\affil[$\dagger$]{Instituto de Matemática Interdisciplinar, Universidad Complutense de Madrid, Plaza de Ciencias 3, E-28040 Madrid, Spain }
  
\affil[$\ddagger$]{ Departamento AGyT, Facultad de Ciencias Matemáticas, Universidad Complutense de Madrid, Plaza de Ciencias 3, E-28040 Madrid, Spain \par\nopagebreak {\small \textit{E-mail}: \texttt{rutwig@ucm.es} } }

\maketitle

\begin{abstract}
\noindent We show that diffeomorphisms of an extended phase space with time, energy, momentum and position degrees of freedom leaving invariant a symplectic $2$-form and a degenerate orthogonal metric ${\rm d} t^2$, corresponding to the Newtonian time line element, locally satisfy Hamilton's equations up to the usual canonical transformation on the position-momentum subspace. 
\end{abstract}

\section{Introduction}

The standard formulation of Hamilton{'}s equations is on a \(2 n\) dimensional symplectic manifold \((\mathbb{P}{}^{\circ},\omega {}^{\circ})\) where
\(\omega {}^{\circ}\) is a closed, non-degenerate 2-form  \cite{ber}. Physically, this is the phase space with local momentum-position coordinates \(y=\{q,p\}\),
\(y\in \mathbb{R}^{2n}\), \(q,p\in \mathbb{R}^n\). { }In these coordinates, the symplectic 2-form locally has the form \(\omega {}^{\circ}=\delta
_{i,j}d p^i\wedge d q^j\), \(i,j,...=1,...,n\). { }Then, the Hamiltonian vector field \(X_{H^{\circ}}\) is defined by (see [2, 3]),
\begin{equation}\label{eqnxref1}
i_{X_{H^{\circ}}}\omega^{\circ}=d H^{\circ},\text{  }H^{\circ}:\mathbb{P}{}^{\circ}\to \mathbb{R}: y\rightarrowtail H^{\circ}(y),
\end{equation}

\noindent where \(H^{\circ}\) is the Hamiltonian that is not time-dependent and the map \(i_X\) maps a vector field \(X\) to a 1-form \(\theta\),
\(X \rightarrowtail \theta =i_X\omega {}^{\circ}=\omega {}^{\circ}(X,\cdot )\). { }In particular this defines an isomorphism between tangent and cotangent
spaces, \(i:T_y\mathbb{P}{}^{\circ}\to T_y^*\mathbb{P}{}^{\circ}:X(y) \rightarrowtail \theta (y)=i_X\omega {}^{\circ}|_y\).

The triple \(\left(\mathbb{P}^{\circ},\omega {}^{\circ},X_{H^{\circ}}\right)\) defines a Hamiltonian system for which the flows \(\phi {}^{\circ}\)
of \(X_H\) satisfy Hamilton{'}s equations,
\begin{equation} \label{eqnxref2}
\phi {}^{\circ}:\mathbb{R}\otimes \mathbb{P}{}^{\circ}\to \mathbb{P}{}^{\circ}: (t,y) \rightarrowtail \phi {}^{\circ}(t,y),\text{  }\frac{d \phi {}^{\circ}(t,y)}{d
t}=X_{H^{\circ}}(\phi^{\circ}(t,y)) .\text{  }
\end{equation}

The symplectomorphisms, or canonical transformations, \(\sigma :\mathbb{P}{}^{\circ}\to \mathbb{P}{}^{\circ}\), leave invariant the 2-form \(\sigma
^*(\omega {}^{\circ})=\omega {}^{\circ}\). The Jacobian matrix of these symplectomorphisms locally take values in the symplectic group, 
\begin{equation}\label{eqnxref3}
\left[\frac{\partial \sigma (y)}{\partial  y}\right]\in \mathcal{S}\mathit{p}(2n),\text{   }y=\{q,p\},\text{  }y\in \mathbb{R}^{2n}, q,p\in \mathbb{R}^n.
\end{equation} 

\subsection{Extended phase space}

This formalism may be generalized to an extended phase space that is a \(2n+2\) dimensional symplectic manifold \((\mathbb{P},\omega )\) with position,
momentum, energy and time coordinates, \(z=(y,\varepsilon ,t)=(q,p,\varepsilon ,t)\) with \(z\in \mathbb{R}^{2n+2}\), \(\varepsilon ,t\in \mathbb{R}\).
In these coordinates, the symplectic metric locally has the form
\begin{equation}
\omega =\delta _{i,j}d p^i\wedge d q^j-d \varepsilon \wedge d t .
\end{equation}

The symplectomorphisms, \(\varrho :\mathbb{P}\to \mathbb{P}\), leave invariant the 2-form \(\varrho ^*(\omega )=\omega\) and their Jacobian matrix
take values in the symplectic group, 
\begin{equation} \label{eqnxref5}
\left[\frac{\partial \varrho (z)}{\partial  z}\right]\in \mathcal{S}\mathit{p}(2n+2),\text{   }z=\{q,p,\varepsilon ,t\}, z\in \mathbb{R}^{2n+2},
\end{equation}

\noindent where \(q,p\in \mathbb{R}^{n}\) and \(\varepsilon ,t\in \mathbb{R}\). { }The extended Hamiltonian function is defined by \(K:\mathbb{P}\to
\mathbb{R}:z\rightarrowtail K(z)\). The triple \(\left(\mathbb{P},\omega ,X_K\right)\) defines a Hamiltonian system for which the flows \(\varphi\) of \(X_K\)
satisfy Hamilton{'}s equations \cite{Lan,Str},
\begin{equation}\label{eqnxref6}
\varphi :\mathbb{R}\otimes \mathbb{P}\to \mathbb{P}: (\tau,z)  \rightarrowtail \varphi (\tau ,z),\text{  }\frac{d \varphi (\tau ,z)}{d \tau }=X_K(\varphi
(\tau ,z)),\text{       }i_{X_K}\omega =d K.
\end{equation}

\noindent \(\tau\) is an invariant parameter and \(\varphi (\tau ,z)=\left(\varphi_q(\tau ,z),  \varphi_p(\tau ,z) ,\varphi _{\varepsilon }(\tau ,z),\varphi _t(\tau
,z)\right)\) where the components are \(\varphi_q(\tau ,z),  \varphi_p(\tau ,z) \in \mathbb{R}^{n}\) and  \(\varphi _{\varepsilon }(\tau ,z)\), \(\varphi
_t(\tau ,z)\in \mathbb{R}\) (\(q,p,\varepsilon\) and \(t\) are just labels of the components). 

In this paper, we are interested in the nonrelativistic case for which the time \(t\) is invariant Newton time.  In the nonrelativistic case, the Hamilton function has the form \(K(q,p,\varepsilon ,t)=H(q,p,t)-\varepsilon \) and
(\ref{eqnxref6}) has locally the form (see equation (11) in \cite{Str}),
\small
\begin{equation}\label{eqnxref7}
\frac{d \varphi (\tau ,z)}{d \tau }= \left(\delta ^{i,j}\left(\frac{\partial H(q,p,t)}{\partial p^i}\frac{\partial }{\partial q^j}-\frac{\partial
H(q,p,t)}{\partial q^i}\frac{\partial }{\partial p^j}\right)+\frac{\partial H(q,p,t)}{\partial t}\frac{\partial }{\partial \varepsilon }+\frac{\partial
}{\partial t}\right)|_{\varphi (\tau ,z)}. 
\end{equation}
\normalsize
The time component of the flows, \(\frac{d \varphi_t (\tau ,z)}{d \tau }=1\) implies \(t=\varphi_t (\tau,z)=\tau\), where the integration constant is set equal to zero. 
Equation (\ref{eqnxref7}) is then locally Hamilton{'}s equations with a Hamiltonian that is explicitly time dependent,
\begin{equation}\label{eqnxref8a}
\frac{d \varphi_q^i (t ,z)}{d t }= \frac{\partial H(q,p,t)}{\partial p^i}, 
\frac{d \varphi_p^i (t ,z)}{d t }= -\frac{\partial H(q,p,t)}{\partial q^i}, 
\frac{d \varphi_{\varepsilon}(t ,z)}{d t }= \frac{\partial H(q,p,t)}{\partial t}.
\end{equation}
The energy component implies \(\varepsilon=\varphi_\varepsilon (t, z)=H(q,p,t)\) and therefore we have the constraint \(K(z)=0\).

If the Hamiltonian is not explicitly time dependent, \(H(q,p,t)\rightarrowtail H^{\circ}(q,p)\)
{ }and if { }\((\mathbb{P}{}^{\circ},\omega {}^{\circ})\) is a symplectic submanifold of \((\mathbb{P},\omega )\), this reduces to the basic formulation
of position-momentum phase space given in equations (\ref{eqnxref1}-\ref{eqnxref3}). 

The above formulations of Hamilton{'}s equations is predicated on the definition of the Hamiltonian function and its associated Hamiltonian vector
fields. This is layered on top of the geometric structure of a symplectic manifold.

\subsection{Jacobi geometry with invariant Newtonian time}

In this paper we describe an alternative formulation of Hamilton's equations on extended phase space \((\mathbb{P},\omega )\) that starts with the
invariance of Newtonian time rather than a Hamiltonian. This is expressed as the geometric structure of a degenerate orthogonal metric \(\lambda\)
and results in the expression of Hamilton{'}s equations in terms of a \textit{Jacobi geometry}, \((\mathbb{P},\omega, \lambda ) \), based on the Jacobi group.

 The alternative formulation requires that the symplectomorphisms (\ref{eqnxref5}) also leave a degenerate orthogonal line element \(\lambda =d\, t^2\) invariant, \(\varrho ^*(\lambda
)=\lambda\). { }This line element defines {`}Newtonian time{'} and we will show that its invariance restricts the Jacobian matrix to be an element
of a subgroup of \(\mathcal{S}\mathit{p}(2n+2)\) that is the Jacobi group, { }\(\mathcal{J}\mathit{a}(n)\equiv \mathcal{H}\mathcal{S}\mathit{p}(2n)\) \cite{eich,ber2,ber3,woit},
\begin{equation}
\left[\frac{\partial \varrho (z)}{\partial z}\right]\in \mathcal{H}\mathcal{S}\mathit{p}(2n)\simeq \mathcal{H}(n)\rtimes \mathcal{S}\mathit{p}(2n)\subset
\mathcal{S}\mathit{p}(2n+2).
\end{equation}

\noindent The Jacobi group \(\mathcal{H}\mathcal{S}\mathit{p}(2n)\) is the semidirect product of a symplectic group \(\mathcal{S}\mathit{p}(2n)\)
and a Weyl-Heisenberg group \(\mathcal{H}(n)\) that is the normal subgroup. This Weyl-Heisenberg group \(\mathcal{H}(n)\) has parameters of velocity
\(v\in \mathbb{R}^n\), force, \(f\in \mathbb{R}^n\) and power \(r\in \mathbb{R}\) (the central coordinate) \cite{foll,Low}. We call these diffeomorphisms
\(\varrho\) with \(\varrho ^*(\omega )=\omega\) and \(\varrho ^*(\lambda )=\lambda\), { } {`}Jacobimorphisms{'}. 

The semidirect product structure \cite{rutwig} of the Jacobi group means that the diffeomorphism \(\varrho\) can always be decomposed into two { }diffeomorphisms
\(\sigma\) and \(\rho\), { }\(\varrho =\rho \circ \sigma\) , \(\tilde{z}=\sigma (z)\).  Therefore,
\begin{equation}
\left[\frac{\partial \varrho (z)}{\partial z}\right]= \left[\frac{\partial \rho \left(\tilde{z}\right)}{\partial \tilde{z}}\right] \left[\frac{\partial
\sigma (z)}{\partial z}\right]=\Upsilon  \Sigma \in \mathcal{H}(n)\rtimes \mathcal{S}\mathit{p}(2n),
\end{equation}

\noindent where
\begin{equation}
\left[\frac{\partial \rho (z)}{\partial z}\right]=\Upsilon (z)\in \mathcal{H}(n), \left[\frac{\partial \sigma (z)}{\partial z}\right]=\Sigma \in
\mathcal{S}\mathit{p}(2n).
\end{equation}

\noindent The \(\sigma\) are just the canonical transformations (\ref{eqnxref5}). We will show that the Jacobimorphisms \(\rho\) {\it locally satisfy
Hamilton{'}s equations.} That is, when the Jacobian is expanded out locally in canonical coordinates, the usual Hamilton{'}s equations result \cite{gold}.

For the Jacobimorphisms \(\rho\), we have specified the invariance of Newtonian time as the starting point. We will show that the Hamiltonian function
and the curves \(\phi\) for the particle trajectories then show up automatically from the functional dependence structure of the Weyl-Heisenberg
subgroup of the Jacobi group. This is spelled out in detail in Section 5 of the paper.

With this context, we are now able to state the key result of this paper as the theorem,

\begin{theorem} Let \textup{ \((\mathbb{P},\omega ,\lambda )\)} be a \(2n+2\) dimensional symplectic manifold \((\mathbb{P},\omega
)\) with position, momentum, energy, time coordinates that has a degenerate orthogonal line element \(\lambda =d\, t^2\) as defined above. { }The
diffeomorphisms, { }\(\varrho :\mathbb{P}\to \mathbb{P}\), that leave \(\omega\) and \(\lambda\) invariant,
\begin{equation}
\varrho ^*(\omega )=\omega , \varrho ^*(\lambda )=\lambda ,
\end{equation}

\noindent are called Jacobimorphisms as their Jacobian matrix takes values in the Jacobi group,
\begin{equation}
\left[\frac{\partial \varrho (z)}{\partial z}\right]=\Gamma (z)\in \mathcal{H}\mathcal{S}\mathit{p}(2n)\simeq \mathcal{H}(n)\rtimes \mathcal{S}\mathit{p}(2n) \subset
\mathcal{S}\mathit{p}(2n+2).
\end{equation}

The semidirect product structure enables the Jacobimorphisms to be written as \textup{ $\varrho $ = $\rho \circ \sigma $} with
\begin{equation}
\left[\frac{\partial \rho (z)}{\partial z}\right]=\Upsilon (z)\in \mathcal{H}(n), \left[\frac{\partial \sigma (z)}{\partial z}\right]=\Sigma (z)\in
\mathcal{S}\mathit{p}(2n),\text{  }\Gamma (z)=\Upsilon (z)\Sigma (z).
\end{equation}

The \(\sigma\) are the usual canonical transformations on position momentum phase space \textup{ \((\mathbb{P}{}^{\circ},\omega {}^{\circ})\)} and
the Jacobian matrix expression for \(\rho\) taking values in the Weyl-Heisenberg group, locally correspond to Hamilton{'}s equations. 

Furthermore, the maximal symmetry group of Hamilton{'}s equations is the group \(\mathcal{H}\mathcal{S}\mathit{p}(2n)\rtimes \mathbb{Z}_2\) where
\(\mathcal{H}\mathcal{S}\mathit{p}(2n)\) is the Jacobi group and \(\mathbb{Z}_2\) is the discrete time reversal symmetry. 
\end{theorem}

The remainder of the paper will prove and explain further this theorem. 

\section{Symplectic 2-form \(\omega\)}

\noindent To provide context and notation for the theorem, consider a \(2n+2\) dimensional { }symplectic manifold \((\mathbb{P},\omega )\). { }\(\mathbb{P}\)
has coordinates \(z:\mathbb{P}\to U\subset \mathbb{R}^{2n+2}\) , \(z=\left\{z^{\alpha }\right\}=\left\{z^1,\text{...}.,z^{2n+2}\right\}\) and \(\alpha
,\beta ,\text{...}=1,\text{...},2n+2\). { }\(\omega\) is a closed non-degenerate two-form. Canonical transformations are a { }symplectomorphism \(\varrho
:\mathbb{P}\to \mathbb{P}\) that leave invariant \(\omega\) in the sense that \(\varrho ^*(\omega )=\omega\). { }Darboux{'}s theorem states that
canonical coordinates exist in a neighbourhood \(U\) of any point \(z\in \mathbb{P}\) so that \(\omega =\zeta _{\alpha ,\beta }d z^{\alpha } d z^{\beta
}\) where the \(2n+2\) dimensional symplectic matrix \(\zeta =\left[\zeta _{\alpha ,\beta }\right]\) has the canonical form,
\begin{equation} \label{eqnxref13}
\zeta =\left[\zeta _{\alpha ,\beta }\right]=\left(
\begin{array}{ccccc}
 0 & 1 & \text{...} & 0 & 0 \\
 -1 & 0 & \text{...} & 0 & 0 \\
 \vdots & \vdots &   & \vdots & \vdots \\
 0 & 0 & \text{...} & 0 & 1 \\
 0 & 0 & \text{...} & -1 & 0 \\
\end{array}
\right)=\left(
\begin{array}{lll}
 \zeta {}^{\circ} & 0 & 0 \\
 0 & 0 & 1 \\
 0 & -1 & 0 \\
\end{array}
\right).
\end{equation}

\noindent (\(\zeta {}^{\circ}\) is a \(2n\) dimensional symplectic submatrix.) Defining \(\tilde{z}=\varrho (z)\), then \(\varrho ^*(\omega )=\omega\)
expressed locally in canonical coordinates as, { }
\begin{equation}
\zeta _{\alpha ,\beta }d \tilde{z}^{\alpha } d \tilde{z}^{\beta }=\zeta _{\alpha ,\beta }\frac{\partial \varrho ^{\alpha }}{\partial z^{\delta
}}\frac{\partial \varrho ^{\beta }}{\partial z^{\gamma }}d z^{\delta }d z^{\gamma }=\zeta _{\alpha ,\beta }d z^{\alpha }d z^{\beta }.
\end{equation}

Defining the Jacobian \(\Sigma _{\delta }^{\alpha }=\frac{\partial \varrho ^{\alpha }}{\partial z^{\delta }}\) and suppressing indices and using
matrix notation \(\Sigma =\left[\frac{\partial \varrho }{\partial z}\right]\), this relation has the form,
\begin{equation}
\varrho ^*(\omega )=d z^{{\rm t} }\Sigma ^{{\rm t}}\zeta  \Sigma  d z=d z^{{\rm t} }\zeta  d z=\omega ,
\end{equation}

\noindent which implies that \(\Sigma ^{{\rm t}}\zeta  \Sigma =\zeta\). { }This is the defining condition for the Jacobian matrices of the symplectomorphism
to be elements of the symplectic group, \(\Sigma \in \mathcal{S}\mathit{p}(2n+2)\).

It is convenient to change notations for the canonical coordinates by using different letters and setting
\begin{equation}
z=\left\{z^{\alpha }\right\}=\left\{z^1,\text{...}.,z^{2n+2}\right\}=\left\{y^1,\text{...}.,y^{2n}, \varepsilon ,t\right\}=\left\{q^1,p^1,\text{...}q^n,p^n,
\varepsilon ,t\right\},
\end{equation}

\noindent and then locally 
\begin{equation}
\omega =\zeta {}^{\circ}{}_{a,b}d y^a d y^b+ d t\wedge d \varepsilon = \delta _{i,j}d p^i\wedge d q^j- d \varepsilon \wedge d t.
\end{equation}

\noindent where \(\zeta {}^{\circ}\) is a \(2n\) dimensional submatrix of \(\zeta\) where \(a,b,\text{...}=1,\text{...}.,2n\) and \(i,j,\text{...}=1,\text{...},n.\)
{ }This notation is chosen as \(z=(q,p,\varepsilon ,t)\) is interpreted physically as the extended phase space degrees of position, momentum, energy
and time degrees of freedom. 

For simplicity, we restrict the symplectic manifold \(\mathbb{P}\) to be of the form \(\mathbb{P}\simeq \mathbb{P}{}^{\circ}\otimes \mathbb{R}^2\)
where \(y\) are canonical coordinates of \(\mathbb{P}{}^{\circ}\) and \((\varepsilon ,t)\) are canonical coordinates of \(\mathbb{R}^2\). { }Clearly
\((\mathbb{P}{}^{\circ},\omega {}^{\circ})\), where
\begin{equation}
\omega {}^{\circ}=\zeta {}^{\circ}{}_{a,b}d y^a d y^b= \delta _{i,j}d p^i\wedge d q^j,
\end{equation}

\noindent is the position-momentum phase space that is a symplectic manifold with symmetry \(\Sigma \in \mathcal{S}\mathit{p}(2n)\).

\section{Degenerate orthogonal metric \(\lambda\)}

Let us now endow \(\mathbb{P}\) with line element that is a degenerate (i.e. singular) orthogonal metric \(\lambda =d t^2\) { }that is defined in
terms of the \(2n+2\) dimensional matrix \(\eta\), 
\begin{equation}\label{eqnxref19}
\lambda =d t^2= \eta _{\alpha ,\beta }d z^{\alpha } d z^{\beta }, \eta =\left[\eta _{\alpha ,\beta }\right]=\left(
\begin{array}{llll}
 0 & \text{...} & 0 & 0 \\
 \vdots &   & \vdots & \vdots \\
 0 & \text{...} & 0 & 0 \\
 0 & \text{...} & 0 & 1 \\
\end{array}
\right).
\end{equation}

We start by considering diffeomorphisms \(\xi :\mathbb{P}\to \mathbb{P}\) that leave only \(\lambda =d t^2\) invariant, \(\xi ^*(\lambda )=\lambda\).
{ }Then locally with \(\tilde{y}=\xi (y),\)
\begin{equation}
\xi ^*(\lambda )=\eta _{\alpha ,\beta }d \tilde{z}^{\alpha } d \tilde{z}^{\beta }=\eta _{\alpha ,\beta }\frac{\partial \xi ^{\alpha }}{\partial
z^{\delta }}\frac{\partial \xi ^{\beta }}{\partial z^{\gamma }}d z^{\delta }d z^{\gamma }=\eta _{\alpha ,\beta }d z^{\alpha }d z^{\beta }=\lambda
.
\end{equation}

\noindent Again, defining the Jacobian \(\Lambda _{\delta }^{\alpha }=\frac{\partial \xi ^{\alpha }}{\partial z^{\delta }}\) and suppressing indices
and using matrix notation, \(\Lambda =\left[\frac{\partial \xi }{\partial z}\right]\), 
\begin{equation}
\xi ^*(\lambda )=d z^{{\rm t} }\Lambda ^{{\rm t}}\eta  \Lambda  d z=d z^{{\rm t} }\eta  d z=\lambda ,
\end{equation}

\noindent and this implies that \(\Lambda ^{{\rm t}}\eta  \Lambda =\eta\). 

\begin{proposition} The matrices \(\Lambda \in \mathcal{G}\mathcal{L}(2n+2,\mathbb{R})\) satisfying \textup{ \(\Lambda ^t\eta  \Lambda
=\eta\)} { }where \(\eta\) is given in (\ref{eqnxref19}) define the extended inhomogeneous general linear matrix group \(\mathcal{I}\mathcal{G}\mathcal{L}(2n+1,\mathbb{R})
\rtimes \mathbb{Z}_2\). 
\end{proposition}

\noindent \textit{ Proof: }Elements \(\Lambda \in \mathcal{G}\mathcal{L}(2n+2,\mathbb{R})\) are realized by nonsingular \(2n+2\) dimensional square
matrices. { }These matrices may be expressed in terms of submatrices as (see e.g. \cite{gil}), 
\begin{equation}
\Lambda =\left(
\begin{array}{ll}
 \Omega  & u \\
 v^{{\rm t}} & \epsilon  \\
\end{array}
\right).\text{   }
\end{equation}

\noindent In this expression, \(\Omega\) is a \(2 n\)+1 dimensional square matrix, \(u,v\in \mathbb{R}^{2n+1}\) are column vectors and \(\epsilon
\in \mathbb{R}\). { } 

The invariance of \(\lambda =d t^2\) is locally defined by \(\Lambda ^{{\rm t}}\eta  \Lambda =\eta\). Substituting (22) and the matrix for \(\eta\) given
in (19) t\textup{ his gives the condition} 
\begin{equation}
\left(
\begin{array}{ll}
 0 & 0 \\
 0 & 1 \\
\end{array}
\right)=\left(
\begin{array}{ll}
 \Omega ^{{\rm t}} & v \\
 u^{{\rm t}} & \epsilon  \\
\end{array}
\right)\left(
\begin{array}{ll}
 0 & 0 \\
 0 & 1 \\
\end{array}
\right)\left(
\begin{array}{ll}
 \Omega  & u \\
 v^{{\rm t}} & \epsilon  \\
\end{array}
\right)=\left(
\begin{array}{ll}
 v v^{{\rm t}} & \epsilon  v \\
 \epsilon  v^{{\rm t}} & \epsilon ^2 \\
\end{array}
\right).
\end{equation}

This implies \(v=0\) and \(\epsilon =\pm 1\) and consequently the invariance of the line element requires \(\Lambda\) to have the form \(\Lambda
=\left(
\begin{array}{ll}
 \Omega  & u \\
 0 & \epsilon  \\
\end{array}
\right)\).

It is convenient to redefine the parameters through the isomorphism, \(u\rightarrowtail \epsilon  u\), so that the matrix factors into the subgroups
\(\Lambda {}^{\circ}\) and \(\Delta\),
\begin{equation}\label{eqnxref24}
\Lambda =\left(
\begin{array}{ll}
 \Omega  & \epsilon  u \\
 0 & \epsilon  \\
\end{array}
\right)=\Lambda {}^{\circ}(\Omega ,u) \Delta (\epsilon )=\left(
\begin{array}{ll}
 \Omega  & u \\
 0 & 1 \\
\end{array}
\right)\left(
\begin{array}{ll}
 1_{2n+1} & 0 \\
 0 & \epsilon  \\
\end{array}
\right).
\end{equation}

\noindent \(\left(1_m\right.\) is notation for a \(m\) dimensional unit matrix.) Note that \(\text{Det} \Delta =\pm 1\) and \(\text{Det} \Lambda
{}^{\circ} \neq 0\) and this implies that Det \(\Omega \neq 0\) and so it is clear that \noindent \(\Lambda {}^{\circ}\in \mathcal{I}\mathcal{G}\mathcal{L}(2n+1,\mathbb{R})\)
{ }and that \(\Delta \in \mathbb{Z}_2\). { }Furthermore, 
\begin{equation}
\Delta (\epsilon ) \Lambda {}^{\circ}(\Omega ,u)\Delta (\epsilon )^{-1}= \Lambda {}^{\circ}(\Omega ,\epsilon  u)\in \mathcal{I}\mathcal{G}\mathcal{L}(2n+1,\mathbb{R}),
\end{equation}

\noindent means that \(\mathcal{I}\mathcal{G}\mathcal{L}(2n+1,\mathbb{R})\) is a normal subgroup of \(\Lambda\). {}Finally, \(\mathcal{I}\mathcal{G}\mathcal{L}(2n+1,\mathbb{R})\cap
\mathbb{Z}_2=e\) and therefore \(\Lambda\) has the semidirect product form, \(\Lambda \in \mathcal{I}\mathcal{G}\mathcal{L}(2n+1,\mathbb{R}) \rtimes
\mathbb{Z}_2\).

Note that \(\mathbb{Z}_2\) is the discrete group of time reversal with elements $\{\pm $1$\}$ such that \(d t\rightarrowtail  d \tilde{t}=\pm  d t\)
and hence \(d \tilde{t}^2= d t^2\). { }Furthermore, the maximal connected subgroup is the inhomogeneous general linear group, { }\(\mathcal{I}\mathcal{G}\mathcal{L}(2n+1,\mathbb{R})\).
{ }

\section{The Jacobi { }group}

We now consider the case \((\mathbb{P},\omega ,\lambda )\) where both the symplectic two-form \(\omega\) and the line element \(\lambda\) are invariants
of the diffeomorphism \(\varrho\), 
\begin{equation}
\varrho ^*(\omega )=\omega , \varrho ^*(\lambda )=\lambda .
\end{equation}

\noindent For both \(\omega\) and \(\lambda\) to be invariant, it is clear from the above that the Jacobian matrices \(\Gamma\) in this case must
satisfy
\begin{equation}
\Gamma ^{{\rm t}}\zeta  \Gamma =\zeta ,\text{  }\Gamma ^{{\rm t}}\eta  \Gamma =\eta .
\end{equation}

\noindent This means that the Jacobian matrices \(\Gamma\) are elements of the group that is the  intersection of the connected component of the inhomogeneous general linear
and symplectic groups. This group is defined to be the \textit{Jacobi} group \(\mathcal{H}\mathcal{S}\mathit{p}(2n)\) \cite{eich,ber2,ber3,woit,Low,Low7} as stated in the following proposition. 

\begin{proposition} { }The intersection of the symplectic groups \(\mathcal{S}\mathit{p}(2n+2)\) and the inhomogeneous
general linear group \textup{ \(\mathcal{I}\mathcal{G}\mathcal{L}(2n+1,\mathbb{R})\)},
\begin{equation}
\mathcal{S}\mathit{p}(2n+2)\cap \mathcal{I}\mathcal{G}\mathcal{L}(2n+1,\mathbb{R})\simeq \mathcal{H}\mathcal{S}\mathit{p}(2n).
\end{equation}
is the Jacobi group \noindent \noindent \(\mathcal{H}\mathcal{S}\mathit{p}(2n)\; \)\noindent  that is the matrix group given by the semidirect product,
\begin{equation}
\mathcal{H}\mathcal{S}\mathit{p}(2n)\simeq \mathcal{H}(n)\rtimes \mathcal{S}\mathit{p}(2n).
\end{equation}
where \textup{\(\mathcal{H}(n)\)} is the Weyl-Heisenberg group. 
\end{proposition}

\noindent \textit{ Proof: }We start by expanding the matrix for the inhomogeneous general linear group defined in (\ref{eqnxref24}),
\begin{equation}
\Omega =\left(
\begin{array}{ll}
 \Sigma  & b \\
 \, c & a \\
\end{array}
\right),\text{      }u=\left(
\begin{array}{l}
 w \\
 2r \\
\end{array}
\right),\text{  }\Lambda {}^{\circ}=\left(
\begin{array}{lll}
 \Sigma  & b & w \\
 \, c & a & 2r \\
 \, 0 & 0 & 1 \\
\end{array}
\right).
\end{equation}

\noindent where { }\(\Sigma\) is an \(n\times n\) matrix, { }\(b,c,w\in \mathbb{R}^n\) and \(a,r\in \mathbb{R}\). The factor of 2 is convenient normalization.
We substitute the symplectic matrix $\zeta $ defined in (\ref{eqnxref13}) and \(\Gamma {}^{\circ}=\Lambda {}^{\circ}\) into the symplectic invariance condition
\(\Gamma {}^{\circ{}{t}}\zeta  \Gamma {}^{\circ}=\zeta\). This requires that
\begin{equation}
\left(
\begin{array}{lll}
 \Sigma ^{{\rm t}}\zeta {}^{\circ}\Sigma  & \Sigma ^{{\rm t}}\zeta {}^{\circ} b & c^{{\rm t}}+\Sigma ^{{\rm t}}\zeta {}^{\circ} w \\
 b^{{\rm t}}\zeta {}^{\circ}\Sigma  & 0 & a +b^{{\rm t}}\zeta {}^{\circ} w \\
 -c+w^{{\rm t}}\zeta {}^{\circ}\Sigma  & \text{  }-a+w^{{\rm t}}\zeta {}^{\circ} b & 0 \\
\end{array}
\right)=\left(
\begin{array}{lll}
 \zeta {}^{\circ} & 0 & 0 \\
 0 & 0 & 1 \\
 0 & -1 & 0 \\
\end{array}
\right).
\end{equation}

\noindent This identity is satisfied with \(b=0\), \(a=1\), \(c=\text{  }w^{{\rm t}}\zeta {}^{\circ} \Sigma\) and \(\Sigma ^{{\rm t}} \zeta {}^{\circ} \Sigma =\zeta
{}^{\circ}\) and therefore,
\begin{equation}
\Gamma {}^{\circ}(\Sigma ,w,r)= \left(
\begin{array}{lll}
 \Sigma  & 0 & w \\
 \, w^{{\rm t}}\zeta {}^{\circ} \Sigma  & 1 & 2r \\
 \, 0 & 0 & 1 \\
\end{array}
\right).
\end{equation}

\noindent It follows that 
\begin{equation} \label{eqnxref34}
\Gamma {}^{\circ}(\Sigma ,w,r)= \Gamma {}^{\circ}\left(1_{2n},w,r\right) \Gamma {}^{\circ}(\Sigma ,0,0)=\Upsilon (w,r)\Sigma  ,\text{  }\Sigma
\in \mathcal{S}\mathit{p}(2n).
\end{equation}

\noindent Furthermore,
\begin{equation}
\Sigma  \Upsilon (w,r)\Sigma ^{-1} =\Upsilon (\Sigma  w,r),
\end{equation}
\begin{equation}
\Gamma {}^{\circ}\left(1_{2n},w,r\right) \cap \Gamma {}^{\circ}(\Sigma ,0,0)=\Upsilon (w,r)\cap  \Sigma =1_{2n}=e .
\end{equation}

\noindent This means that \(\Upsilon (w,r)\in \mathcal{H}(n)\) is a normal subgroup of \(\mathcal{H}\mathcal{S}\mathit{p}(2n)\) { }and the conditions
are satisfied for the Jacobi group to have the semidirect product structure claimed in Proposition 3. 

\subsection{Weyl-Heisenberg group}

The final step is to show that the \(\Upsilon (w,r)\) are elements of { }the Weyl-Heisenberg group \(\mathcal{H}(n)\). { }The elements \(\Upsilon
(w,r)\) define a matrix group with group multiplication and inverse given by
\begin{equation}\label{eqnxref37}
\Upsilon \left(w',r'\right)\Upsilon (w,r)=\Upsilon \left(w'+w,r'+r+\, \frac{1}{2}w'^t\zeta {}^{\circ} w\right),
\end{equation}
\begin{equation}
\Upsilon ^{-1}(w,r)=\Gamma (-w,-r),\; e=\Upsilon (0,0).
\end{equation}

These are the group product and inverse for elements of the Weyl-Heisenberg group, \(\Upsilon (w,r)\in \mathcal{H}(n)\), { }in unpolarized coordinates
\cite{foll,Low,Low7}. Computing the Lie algebra of \(\mathcal{H}(n)\) is the Weyl-Heisenberg group gives the expected result. { }A basis of the matrix Lie
algebra is,
\begin{equation}
W_a=\frac{\partial }{\partial  w^a}\Upsilon (w,r)|_{w=r=0}, R=\frac{\partial }{\partial  r}\Upsilon (w,r)|_{w=r=0}.
\end{equation}

\noindent A generic element of the algebra is { }\(Z=w^{\alpha }W_{\alpha }+r R\). As the Lie algebra of a matrix group is the usual matrix commutator
\([A,B]=A B-B A\), the following relations follow at once,
\begin{equation}
\left[W_{\alpha },W_{\beta }\right]=\zeta {}^{\circ}{}_{\alpha ,\beta }R,\text{   }\left[W_{\alpha },R\right]=0.
\end{equation}

\noindent These are easily seen to correspond to the Weyl-Heisenberg algebra, where \(R\) is the central generator. { }

\subsection{Jacobi group}

The elements \(\Gamma {}^{\circ}(\Sigma ,w,r)=\Upsilon (w,r)\Sigma\) define the Jacobi matrix group \(\mathcal{H}\mathcal{S}\mathit{p}(2n)\) with
group product,
\begin{equation}
 \Gamma {}^{\circ}\left(\Sigma ',w',r'\right)\Gamma {}^{\circ}(\Sigma ,w,r) =\Gamma {}^{\circ}\left(\Sigma '\Sigma ,w'+\Sigma 'w,r'+r+\, \frac{1}{2}\,
w'^{{\rm t}}\zeta {}^{\circ} \Sigma ' w\right),
\end{equation}

\noindent and inverse
\begin{equation}
 \Gamma {}^{\circ}(\Sigma ,w,r)^{-1} =\Gamma {}^{\circ}\left( \Sigma ^{-1},-\Sigma ^{-1}w,-r\right).
\end{equation}

\noindent This completes the proof of Proposition 3. { }

\begin{proposition} { }The intersection of the symplectic group, \(\mathcal{S}\mathit{p}(2n+2)\) and extended inhomogeneous
general linear group, \textup{ \(\mathcal{I}\mathcal{G}\mathcal{L}(2n+1,\mathbb{R})\rtimes \mathbb{Z}_2\)}\textup{  }is \(\mathcal{H}\mathcal{S}\mathit{p}(2n)\rtimes
\mathbb{Z}_2\),
\begin{equation}
\mathcal{S}\mathit{p}(2n+2)\cap \mathcal{I}\mathcal{G}\mathcal{L}(2n+1,\mathbb{R})\rtimes \mathbb{Z}_2\simeq \mathcal{H}\mathcal{S}\mathit{p}(2n)\rtimes
\mathbb{Z}_2.
\end{equation}
\end{proposition}
\noindent \textit{ Proof: }Elements of the full group including the discrete \(\mathbb{Z}_2\) time reversal symmetry are defined by \(\Gamma (\Sigma
,w,r,\epsilon )=\Gamma {}^{\circ}(w,r)\Delta (\epsilon )\) and it is straightforward to verify that this matrix group coincides with the semidirect
product \(\mathcal{H}\mathcal{S}\mathit{p}(2n)\rtimes \mathbb{Z}_2\). { }The automorphisms are
\begin{equation}
\Delta (\epsilon ) \Gamma {}^{\circ}(\Sigma ,w,r)\Delta (\epsilon )^{-1} =\Gamma {}^{\circ}(\Sigma ,\epsilon  w, r).
\end{equation}

\noindent \(\mathcal{H}\mathcal{S}\mathit{p}(2n)\) is therefore a normal subgroup of the full group and \(\mathbb{Z}_2\cap \mathcal{H}\mathcal{S}\mathit{p}(2n)=e\).
The { }semidirect product structure stated in Proposition 4 follows directly. Consequently, \(\mathcal{H}\mathcal{S}\mathit{p}(2n)\rtimes \mathbb{Z}_2\) is the maximal group that leaves the symplectic 2-form
\(\omega\) and Newtonian time line element \(\lambda\) invariant.

\subsection{Jacobimorphisms}

\begin{corollary} { }Let \(\varrho\) be a symplectomorphism \(\varrho :\mathbb{P}\to \mathbb{P}\), \textup{ \(\varrho ^*(\omega
)=\omega\)}, preserving the Newtonian time line element, \textup{ \(\varrho ^*(\lambda )=\lambda\)}. { }Then, locally the Jacobian matrix of \(\varrho\)
is an element of the Jacobi group, \textup{ \(\mathcal{H}\mathcal{S}\mathit{p}(2n)\)}. { }\textup{ (}Recall that these are called Jacobimorphisms.\textup{
)}
\end{corollary}

\noindent \textit{ Proof}:  The entries of the Jacobian matrix \(\left[\frac{\partial \varrho }{\partial z}\right]\) are continuous functions. { }Therefore, the Jacobian matrix must be an element of the connected component of \(\mathcal{H}\mathcal{S}\mathit{p}(2n)\rtimes \mathbb{Z}_2\). { }The connected component
is the Jacobi group, \(\mathcal{H}\mathcal{S}\mathit{p}(2n)\) and therefore

\begin{equation}
 \left[\frac{\partial \varrho }{\partial z}\right]\in \mathcal{H}\mathcal{S}\mathit{p}(2n).
\end{equation}

\section{Hamilton's equations}

\noindent We can use the property of the semidirect product form of the Jacobi group to expand this expression for the Jacobian matrix. Set { }\(\varrho =\rho \circ \sigma\)
where,

\begin{equation}
\rho :\mathbb{P}\to \mathbb{P}:z'\rightarrowtail z^{\prime\prime }=\rho \left(z'\right), \sigma :\mathbb{P}\to \mathbb{P}:z\rightarrowtail z'=\sigma
(z)
\end{equation}

\noindent are diffeomorphisms. Then, the Jacobian is 
\begin{equation}
\left[\frac{\partial \varrho (z)}{\partial z}\right]=\left[\frac{\partial \rho \left(z'\right)}{\partial
z'}\right]\left[\frac{\partial \sigma (z)}{\partial z}\right].
\end{equation}

\noindent Using the semidirect product factorization (\ref{eqnxref34}), { }
\begin{equation}
\left[\frac{\partial \rho (z)}{\partial z}\right]=\Upsilon (w(z),r(z))\in \mathcal{H}(n),\text{   }
\end{equation}
\begin{equation}
\left[\frac{\partial \sigma (z)}{\partial z}\right]=\Gamma (\Sigma (z),0,0)\simeq \Sigma (z)\in \mathcal{S}\mathit{p}(2 n).
\end{equation}

\subsection{Canonical transformations}

Expand \(z\) { }with the component notation \(\{z\}=\{y,\varepsilon ,t\}\) with \(z\in \mathbb{R}^{2n+2}\), \(y\in \mathbb{R}^{2n}\) and \(\varepsilon
,t\in \mathbb{R}\), and likewise use these to label the components of the diffeomorphism $\sigma $,
\begin{equation}
\{\sigma (z)\} =\left\{\sigma _y(y,\varepsilon ,t),\sigma _{\varepsilon }(y,\varepsilon ,t),\sigma _t(y,\varepsilon ,t)\right\}.
\end{equation}

\noindent The Jacobian matrix may be expanded explicitly into the \(2n+2\) dimensional matrix, { }
\begin{equation}\label{eqnxref51}
\left(
\begin{array}{lll}
 \frac{\partial \sigma _y(y, \varepsilon , t)}{\partial y} & \frac{\partial \sigma _y(y, \varepsilon , t)}{\partial \varepsilon } & \frac{\partial
\sigma _y(y, \varepsilon , t)}{\partial t} \\
 \frac{\partial \sigma _{\varepsilon }(y, \varepsilon , t)}{\partial y} & \frac{\partial \sigma _{\varepsilon }(y, \varepsilon , t)}{\partial \varepsilon
} & \frac{\partial \sigma _{\varepsilon }(y, \varepsilon , t)}{\partial t} \\
 \frac{\partial \sigma _t(y, \varepsilon , t)}{\partial y} & \frac{\partial \sigma _t(y, \varepsilon , t)}{\partial \varepsilon } & \frac{\partial
\sigma _t(y, \varepsilon , t)}{\partial t} \\
\end{array}
\right)=\left(
\begin{array}{lll}
 \Sigma (z) & 0 &  0 \\
 \, 0  & 1 & 0 \\
 \, 0 & 0 & 1 \\
\end{array}
\right),
\end{equation}

\noindent where we are suppressing indices and using matrix notation for the Jacobian submatrices. Note that { }\(\Sigma (z)\) { }is an \(2n\times
2n\) submatrix. This restricts the functional form of \(\sigma\) so that the solution is just the canonical transformation. { }That is,
\begin{equation}
\frac{\partial \sigma _y(y, \varepsilon , t)}{\partial \varepsilon }=0, \frac{\partial \sigma _y(y, \varepsilon , t)}{\partial t}=0.
\end{equation}

\noindent implies \(\sigma _y(y,\varepsilon ,t)=\sigma _y(y)\equiv \sigma (y)\) and likewise for \(\sigma _{\varepsilon }\) and \(\sigma _t\), where
we set integration constants to zero, yielding,
\begin{equation}
\begin{array}{l}
 \tilde{y}=\sigma _y(y,\varepsilon ,t)=\sigma (y),\text{  }\frac{\partial \sigma (y)}{\partial y}=\Sigma (y), \\
 \tilde{\varepsilon }=\sigma _{\varepsilon }(y,\varepsilon ,t)=\varepsilon , \\
 \tilde{t}=\sigma _t(y,\varepsilon ,t)= t.\\
\end{array}
\end{equation}

\noindent The diffeomorphisms \(\sigma {}^{\circ}:\mathbb{P}{}^{\circ}\to \mathbb{P}{}^{\circ}\) are precisely the canonical transformations on \(\mathbb{P}{}^{\circ}\).

\subsection{Hamilton{'}s equations }

The diffeomorphisms \(\rho (z)\) may likewise be expanded as 
\begin{equation}
\{\rho (z)\} =\left\{\rho _y(y,\varepsilon ,t),\rho _{\varepsilon }(y,\varepsilon ,t),\rho _t(y,\varepsilon ,t)\right\}.
\end{equation}

\noindent and the \(2n+2\) dimensional Jacobian matrix is,
\begin{equation}
\left(
\begin{array}{lll}
 \frac{\partial \rho _y(y, \varepsilon , t)}{\partial y} & \frac{\partial \rho _y(y, \varepsilon , t)}{\partial \varepsilon } & \frac{\partial \rho
_y(y, \varepsilon , t)}{\partial t} \\
 \frac{\partial \rho _{\varepsilon }(y, \varepsilon , t)}{\partial y} & \frac{\partial \rho _{\varepsilon }(y, \varepsilon , t)}{\partial \varepsilon
} & \frac{\partial \rho _{\varepsilon }(y, \varepsilon , t)}{\partial t} \\
 \frac{\partial \rho _t(y, \varepsilon , t)}{\partial y} & \frac{\partial \rho _t(y, \varepsilon , t)}{\partial \varepsilon } & \frac{\partial \rho
_t(y, \varepsilon , t)}{\partial t} \\
\end{array}
\right)=\left(
\begin{array}{lll}
 1_{2n} & 0 &  w(z) \\
 \, w^t(z)\zeta {}^{\circ}  & 1 & 2 r(z) \\
 \, 0 & 0 & 1 \\
\end{array}
\right).
\end{equation}

This restricts the functional dependency of the diffeomorphisms \(\rho\) as follows. { }First the time component,
\begin{equation}
\frac{\partial \rho _t(y, \varepsilon , t)}{\partial y}=0, \frac{\partial \rho _t(y, \varepsilon , t)}{\partial \varepsilon }=0, \frac{\partial
\rho _t(y, \varepsilon , t)}{\partial t}=1,
\end{equation}

\noindent and so setting the integration constant equal to zero results in \(\rho _t(y,\varepsilon ,t)=t.\) Next for the energy component, note that
the only functional dependency constraint is, 
\begin{equation}
\frac{\partial \rho _{\varepsilon }(y, \varepsilon , t)}{\partial \varepsilon }=1.
\end{equation}

\noindent Therefore \(\rho _{\varepsilon }\) may be written as \(\rho _{\varepsilon }(y,\varepsilon ,t)=\varepsilon +H(y,t)\) where \(H\) is some
function \(H:\mathbb{P}{}^{\circ}\otimes \mathbb{R}\to \mathbb{R}\). { }

Finally, for the \(\rho _y\) components, 
\begin{equation}
\frac{\partial \rho _y(y,\varepsilon ,t)}{\partial \varepsilon }=0 ,\text{  }
\end{equation}

\noindent and consequently \(\rho _y(y,\varepsilon ,t)=y+ \phi (t)\) where { }\(\phi\) is some function, \(\phi :\mathbb{R}\to \mathbb{P}{}^{\circ}\).

Summarizing, due to the functional dependency constraints, the diffeomorphism \(\tilde{z}=\rho (z)\) is expanded as 
\begin{equation}
\begin{array}{l}
 \tilde{y}=\rho _y(y,\varepsilon ,t)=\phi (t,y)=y+ \phi (t), \\
 \tilde{\varepsilon }=\rho _{\varepsilon }(y,\varepsilon ,t)=\varepsilon +H(y,t), \\
 \tilde{t}=\rho _t(y,\varepsilon ,t)= t. \\
\end{array}
\end{equation}

\noindent \(H\) and \(\phi\) are functions that are not specified \textit{ a priori, } but rather arise naturally as a result of the functional dependence
constraints that the Weyl-Heisenberg subgroup of the Jacobi group place on the Jacobians \(\frac{\partial \rho (z)}{\partial z}\),
\begin{equation}
\begin{array}{l}
 H: \mathbb{P}{}^{\circ}\otimes \mathbb{R}\to \mathbb{R}:(y,t)\rightarrowtail H(y,t), \\
 \phi : \mathbb{R}\otimes \mathbb{P}{}^{\circ}\to \mathbb{P}{}^{\circ}:t\rightarrowtail \phi (t,y). \\
\end{array}
\end{equation}

\noindent \(H\) will turn out to be the Hamiltonian and \(\phi _y\) correspond to the curves that are the trajectories in phase space that are solutions
to Hamilton{'}s equations.

Substituting these back into (\ref{eqnxref51}), the Jacobian matrix now has the form
\begin{equation}
\left(
\begin{array}{lll}
 1_{2n} & 0 & \frac{d \phi (t, y)}{d t} \\
 \frac{\partial H(y, t)}{\partial y} & 1 & \frac{\partial H(y, t)}{\partial t} \\
 0 & 0 & 1 \\
\end{array}
\right)=\left(
\begin{array}{lll}
 1_{2n} & 0 &  w(y,t) \\
 \, w^t(y,t)\zeta {}^{\circ}\text{   } & 1 & 2r(y,t) \\
 \, 0 & 0 & 1 \\
\end{array}
\right).
\end{equation}

\noindent Therefore we have
\begin{equation}
\frac{d \phi (t,y)}{d t}=w(y,t),\text{  }\frac{\partial H(y,t)}{\partial y}=w^t(y,t)\zeta {}^{\circ},\text{   }\frac{\partial H(y,t)}{\partial
t}=2 r(y,t).
\end{equation}

\noindent Re-arranging the above expression results in,
\begin{equation}
\text{}\frac{d \phi (t,y)}{d t}=w(y,t)=\zeta {}^{\circ}\left[\frac{\partial H(y,t)}{\partial y}\right]^t,\text{   }\frac{\partial H(y,t)}{\partial
t}=2r(y,t).
\end{equation}

\noindent In component form, these expressions adopt the form, 
\begin{equation}\label{eqnxref64}
\text{}\frac{d \phi ^{\alpha }(t,y)}{d t}=\zeta {}^{\circ{}{\alpha,\beta}}\frac{\partial H(y,t)}{\partial y^{\beta }},\text{   }\frac{\partial
H(y,t)}{\partial t}=2r(y,t),
\end{equation}

\noindent where \(\left[\zeta {}^{\circ{}{\alpha,\beta}}\right]=-\zeta {}^{\circ}\). { }These are easily recognized as Hamilton's equations. 

The full solution is the composition { }\(\varrho =\rho \circ \sigma\). { }This is just a canonical transformation on \(\mathbb{P}{}^{\circ}\). In
terms of the components, this is
\begin{equation}
\begin{array}{l}
 \tilde{y}=\varrho _y(y,\varepsilon ,t)=\sigma (y)+ \phi (t), \\
 \tilde{\varepsilon }=\varrho _{\varepsilon }(y,\varepsilon ,t)=\varepsilon +H(\sigma (y),t), \\
 \tilde{t}=\varrho _t(y,\varepsilon ,t)= t. \\
\end{array}
\end{equation}

\noindent These satisfy Hamilton{'}s equations as shown above. 

The converse requires us to prove that, if the diffeomorphisms satisfy Hamilton's equations (\ref{eqnxref64}), then the symplectic 2-form \(\omega\) and line
element \(\lambda\) are invariant. Suppressing indices and using matrix notation, { } 
\begin{equation}
\begin{array}{ll}
 \tilde{\omega } & =d \tilde{y}^t\zeta {}^{\circ} d \tilde{y} +d\tilde{t}\wedge d \tilde{\varepsilon } \\
   & =(d y+d \phi (t))^t\zeta {}^{\circ} (d y+d \phi (t))+ d t\wedge (d \varepsilon +d H(y,t)) \\
   & =d y^t\zeta {}^{\circ} d y +d t\wedge d \varepsilon +\left[\frac{d \phi (t)}{d t}\right]^t \zeta {}^{\circ} d y\wedge d t-\text{  }\frac{\partial
 H(y,t)}{d y}d y\wedge d t \\
   & =\omega - \left(\left[\zeta {}^{\circ}\frac{d \phi (t)}{d t}\right]^t +\frac{\partial  H(y,t)}{d y}\right)d y\wedge d t=\omega . \\
\end{array}
\end{equation}

\noindent \(\lambda =d t^2\) is invariant as \(t\) is an invariant parameter in Hamilton's equations. { }

\noindent This completes the proof of {\it Theorem }1.

\section{Discussion of the theorem}

\noindent The symplectic symmetry and inhomogeneous general linear symmetries are very well known fundamental symmetries of classical mechanics.
{ }It should not therefore be a surprise that the intersection of these symmetries, where both are manifest, plays a fundamental role in Hamilton{'}s
mechanics. 

The symplectic transformations \(\sigma\) on the position-momentum subspace \((\mathbb{P}{}^{\circ},\omega {}^{\circ})\) with the Jacobian matrix
\(\left[\frac{\partial \sigma (z)}{\partial z}\right]=\Gamma (\Sigma (z),0,0)\simeq \Sigma (z)\in \mathcal{S}\mathit{p}(2n)\) are the usual canonical
transformations of basic Hamilton{'}s mechanics. These are well known and have been extensively studied (see e.g. \cite{gold}).

It is the appearance of the Weyl-Heisenberg group that is novel and perhaps unexpected as it is generally associated with quantum mechanics. But it
is just the semidirect product of two abelian groups and appears in many other contexts; this is just one of those cases. Further expanding the
coordinates \(y\rightarrowtail (q,p)\), we discuss the physical meaning of the Weyl-Heisenberg group in Hamilton{'}s
mechanics. 

\smallskip
Define \(y=(q,p)\), \(q,p\in \mathbb{R}^n\) { }and \(\phi =(\pi ,\xi )\), \(\pi(t) ,\xi(t)\in \mathbb{R}^n\).  As is usual, \(p\) is canonical momentum and \(q\) is canonical
position. We also scale \(2 r\rightarrowtail r\). We will continue to use matrix notation with indices suppressed. { }Hamilton's equations then take
on their most simple form,
\begin{equation}\label{eqnxref67}
\text{}\frac{d \xi (t)}{d t}=v=\frac{\partial H(q,p,t)}{\partial p},\text{}\frac{d \pi (t)}{d t}=f=-\frac{\partial H(q,p,t)}{\partial q},\text{
  }\frac{\partial H(q,p,t)}{\partial t}=r,
\end{equation}

\noindent where \(v(q,p,t),f(q,p,t)\in \mathbb{R}^n\) are the velocity and force respectively and \(r(q,p,t)\in \mathbb{R}\) is the power. That
the velocity, force and power are functions of \((q,p,t)\) is implicit in the above equation and will be implicit in the following. The Weyl-Heisenberg subgroup in these coordinates
is \(\Upsilon (v,f,r)=\Gamma \left(1_{2n},v,f,r\right)\) and equation (\ref{eqnxref5}) expands as
\begin{equation}
\left(
\begin{array}{llll}
 1_n & 0 & 0 & v \\
 0 & 1_n & 0 & f \\
 -f^t & v^t & 1 & r \\
 0 & 0 & 0 & 1 \\
\end{array}
\right)=\left(
\begin{array}{llll}
 1_n & 0 & 0 & \frac{d \xi (t)}{d t} \\
 0 & 1_n & 0 & \frac{d \pi (t)}{d t} \\
 \left(\frac{\partial H(q,p,t)}{\partial q}\right)^t & \left(\frac{\partial H(q,p,t)}{\partial p}\right)^t & 1 & \frac{\partial H(q,p,t)}{\partial
t} \\
 0 & 0 & 0 & 1 \\
\end{array}
\right).
\end{equation}

The coordinates \(z\) of the extended phase space \(\mathbb{P}\) may be similarly expanded as \(z= (p,q,\varepsilon  ,t)\) and the Weyl-Heisenberg
transformation \(d \tilde{z}=\Upsilon  d z\) expands as 
\begin{equation}\label{eqnxref69}
\left(
\begin{array}{l}
 d \tilde{q} \\
 d \tilde{p} \\
 d \tilde{\varepsilon } \\
 d \tilde{t} \\
\end{array}
\right)=\left(
\begin{array}{llll}
 1_n & 0 & 0 & v \\
 0 & 1_n & 0 & f \\
 -f^t & v^t & 1 & r \\
 0 & 0 & 0 & 1 \\
\end{array}
\right)\left(
\begin{array}{l}
 d q \\
 d p \\
 d \varepsilon  \\
 d t \\
\end{array}
\right).
\end{equation}

\noindent Using Hamilton's equations (\ref{eqnxref67}), this results in
\begin{equation}\label{eqnxref70}
\begin{array}{ll}
 d\tilde{t}=d t, &   \\
 d\tilde{q}=d q+v d t & \text{\textit{$=$}}\text{\textit{$ $}}\text{\textit{$d q+d $}}\xi \text{\textit{$(t),$}} \\
 d\tilde{p}=d p+f d t & \text{\textit{$=$}}\text{\textit{$ $}}\text{\textit{$d p+d $}}\pi \text{\textit{$(t),$}} \\
 d \tilde{\varepsilon }= d \varepsilon +v \cdot d p-f\cdot d q+r d t & = d \varepsilon +d H(p,q,t). \\
\end{array}
\end{equation}

These are the expected transformations that relate two states in extended phase space that have a relative rate of change of position, momentum and
energy with respect to time. That is, they have a relative velocity \(v\), force \(f\) and power \(r\). { }These are general states in the extended
phase space that may be inertial or noninertial. { }In the energy transformation, \(\int v\cdot d p\) is the incremental kinetic energy and \(-\int
f\cdot d q\) is the work transforming from energy state \(\varepsilon\) to \(\tilde{\varepsilon  }\). The term \(\int r d t\) is the energy for the
power \(r\) for time dependent Hamiltonians. { }Solving Hamilton's equations enables these to be integrated to the form,
\begin{equation}
\begin{array}{l}
 \tilde{t}=\rho _t(t) =t, \\
 \tilde{q}=\rho _q(q,t)=q + \xi (t), \\
 \tilde{p}=\rho _p(p,t)=p + \pi (t),  \\
 \tilde{\varepsilon } =\rho _{\varepsilon }(\varepsilon ,q,p,t)= \varepsilon  + H(q,p,t). \\
\end{array}
\end{equation}

Using the group multiplication (\ref{eqnxref37}) with \(\Sigma =1_{2n}\) , or simply multiplying the matrices in (\ref{eqnxref69}) together shows that { }
\begin{equation}
\Upsilon \left(\tilde{v},\tilde{f},\tilde{r}\right)\Upsilon (v,f,r)=\Upsilon \left(v+\tilde{v},f+\tilde{f},r+\tilde{r}+\tilde{f} v-\tilde{v} f\right),
\end{equation}
\begin{equation}
\Upsilon (v,f,r)\Upsilon \left(\tilde{v},\tilde{f},\tilde{r}\right)=\Upsilon \left(v+\tilde{v},f+\tilde{f},r+\tilde{r}-\tilde{f} v+\tilde{v} f\right).
\end{equation}

Consequently, two noninertial transformations do not commute, as expected.  The noncommutativity is fully determined by the Weyl-Heisenberg group.

\subsection{Inertial case}

The particular case where the force \(f=0\) and power \(r=0\) corresponds to inertial transformations. { }This defines an abelian subgroup \(\mathcal{A}(n)\simeq
\left(\mathbb{R}^n,+\right)\) of the Weyl-Heisenberg group, \(\mathcal{A}(n)\subset \mathcal{H}(n)\) with elements \(\Upsilon\) satisfying the group
product and inverse relations, 
\begin{equation}
\Upsilon \left(v',0,0\right)\Upsilon (v,0,0)=\Upsilon \left(v'+v,0,0\right), \Upsilon ^{-1}(v,0,0)=\Upsilon (-v,0,0).
\end{equation}

A particular subgroup of the symplectic group is the special orthogonal group, \(\mathcal{S}\mathcal{O}(n)\subset \mathcal{S}\mathit{p}(2n)\), with
elements \(\Sigma\) restricted to the rotations
\begin{equation}
\Sigma (R)=\left(
\begin{array}{ll}
 R & 0 \\
 0 & R \\
\end{array}
\right), R\in \mathcal{S}\mathcal{O}(n), \Sigma \in \mathcal{S}\mathcal{O}(n)\subset \mathcal{S}\mathit{p}(2n).
\end{equation}

The Euclidean group on phase space is the semidirect product \(\mathcal{E}(n)\simeq \mathcal{A}(n)\rtimes \mathcal{S}\mathcal{O}(n)\) with
elements \(\Gamma (\Sigma (R),v,0,0)=\Upsilon (v,0,0)\Sigma (R)\) that is clearly a subgroup of the Jacobi group. { }The resulting transformations
are the special case of (\ref{eqnxref70}) that are the expected inertial case, 
\begin{equation}
\begin{array}{l}
 d\tilde{t}=d t, \\
 d\tilde{q}=R d q + v d t, \\
 d\tilde{p}=R d p,  \\
 d \tilde{\varepsilon } = d \varepsilon  + v \cdot d p.  \\
\end{array}
\end{equation}

We emphasize that the Euclidean group \(\mathcal{E}(n)\) is \textit{not} a subgroup of the symplectic group \(\mathcal{S}\mathit{p}(2n)\)  but it is a subgroup of the Jacobi group, { }\(\mathcal{E}(n)\subset \mathcal{H}\mathcal{S}\mathit{p}(2n)\simeq \mathcal{H}(n)\rtimes
\mathcal{S}\mathit{p}(2n)\).

\section{Conclusion}

In this paper, we have considered an alternative approach to the extended Hamiltonian formalism \cite{Lan,Str} that exploits the group theoretical formalism, instead of the usual analytic approach. 

It is generally asserted that the symplectic group \(\mathcal{S}\mathit{p}(2n)\) together with the time reversal symmetry \(\mathbb{Z}_2\) is the
maximal symmetry of Hamilton's equations (see \cite{arn,abra, gold}). { }We have shown that the Euclidean group \(\mathcal{E}(n)\) \textit{ is not} a subgroup
of \(\mathcal{S}\mathit{p}(2n)\rtimes \mathbb{Z}_2\). { }This Euclidean group is the homogeneous subgroup of Galilean relativity. Hamilton{'}s
equations are a formulation of nonrelativistic (i.e. Galilean relativistic) mechanics and so must have the Euclidean group as a symmetry for the
inertial case. { }Consequently the assertion that \(\mathcal{S}\mathit{p}(2n)\rtimes \mathbb{Z}_2\) is the maximal symmetry group of Hamilton{'}s
equations \textit{ cannot be correct}. 

We have shown that Hamilton{'}s equations can be expressed in terms of Jacobimorphisms that are natural geometric constructs defined only by the
invariance of a symplectic 2-form and a degenerate orthogonal metric representing Newtonian time. An immediate consequence of this is that the maximal
symmetry of Hamilton{'}s equations is the extended Jacobi group,
\begin{equation}
\mathcal{H}\mathcal{S}\mathit{p}(2n)\rtimes \mathbb{Z}_2\simeq \mathcal{H}(n)\rtimes \mathcal{S}\mathit{p}(2n)\rtimes \mathbb{Z}_2.
\end{equation}

We have further shown above that the Jacobi group contains the Euclidean group as a subgroup, \(\mathcal{E}(n)\subset \mathcal{H}\mathcal{S}\mathit{p}(2n)\),
for the inertial case as required by Galilean relativity. 

We note { }that the same methodology may be applied to the special relativistic case \((\mathbb{P},\omega ,\lambda )\) where invariant time is defined
by Einstein proper time that is also a degenerate orthogonal metric on extended phase space,
\begin{equation}
\lambda = d\, \tau ^2= d\, t^2 - \frac{1}{c^2}d\, q^2.
\end{equation}

One obtains a corresponding symmetry group that results in the expected relativistic generalization of Hamilton{'}s equations \cite{Low6}. These equations
contract to the expected nonrelativistic results in the limit \(c\to \infty\). { }This motivates a study of invariant time based on a nondegenerate
Born metric on extended phase for which interesting new physics results \cite{Low8}.

\section*{Acknowledgements}

 \small
 
R.C.S. has been partially supported by Agencia Estatal de Investigaci\'on (Spain) under grant  PID2019-106802GB-I00/AEI/10.13039/501100011033.

\end{document}